\documentclass{jpp}
\usepackage{graphicx}

\usepackage[utf8]{inputenc}
\usepackage[T1]{fontenc}
\usepackage{amsmath}
\usepackage{qcircuit}
\usepackage{tikz}

\newcommand{\no}{\nonumber}
\newcommand{\tanc}{| \mathrm{ancilla} \rangle}
\newcommand{\tphys}{| \mathrm{phys} \rangle}
\newcommand{\tx}{| \mathrm{phys;} x \rangle}

\newcommand{\tvx}{| \mathrm{phys;}v_x \rangle}

\shorttitle{Hamiltonian Simulations for the Vlasov-Maxwell equations}
\shortauthor{H. Higuchi and others}

\title{Quantum Calculation for Two-Stream Instability and Advection Test of Vlasov-Maxwell Equations: Numerical Evaluation of Hamiltonian Simulation
}

\author{Hayato Higuchi\aff{1}
  \corresp{\email{higuchi.hayato.007@gmail.com}},
  J. W. Pedersen\aff{2},
  K. Toyoizumi\aff{3},
  K. Yoshikawa\aff{4},
  C. Kiumi\aff{5},
 \and A. Yoshikawa\aff{1}}

\affiliation{
\aff{1}Graduate School of Science, Kyushu University, Motooka, Nishi-ku, Fukuoka 819-0395, Japan
\aff{2}RIKEN Center for Quantum Computing, Wako, Saitama, 351-0198, Japan
\aff{3}Graduate School of Science and Technology, Keio University, Hiyoshi 3-14-1, Kohoku-ku, Yokohama, 223-8522, Japan
\aff{4}Center for Computational Sciences, University of Tsukuba, 1-1-1 Tennodai, Tsukuba 305-8577, Japan
\aff{5}Center for Quantum Information and Quantum Biology, Osaka University,
1-2 Machikaneyama, Toyonaka 560-0043, Japan
}

\begin{document}

\maketitle

\begin{abstract}
The Vlasov-Maxwell equations provide kinetic simulations of collisionless plasmas, but numerically solving them on classical computers is often impractical.
This is due to the computational resource constraints imposed by the time evolution in the 6-dimensional phase space, which requires broad spatial and temporal scales.
In this study, we develop a quantum-classical hybrid Vlasov-Maxwell solver.
Specifically, the Vlasov solver implements the Hamiltonian simulation based on Quantum Singular Value Transformation~(QSVT), coupled with a classical Maxwell solver.

We perform numerical simulation of a 1D advection test and a 1D1V two-stream instability test on the \texttt{Qiskit-Aer-GPU} quantum circuit emulator with an A100 GPU.
The computational complexity of our quantum algorithm can potentially be reduced from the classical $O(N^6T^2)$ to $O(\text{poly}(\log(N),N,T))$ for the $N$ grid system and simulation time $T$.
Furthermore, the numerical analysis reveals that our quantum algorithm is robust under larger time steps compared with classical algorithms with the constraint of Courant-Friedrichs-Lewy~(CFL) condition.

\end{abstract}

\section{Introduction}
Plasma simulations have historically utilized supercomputers to accurately predict and understand the behavior of electromagnetic plasmas in large-scale astrophysical phenomena, nuclear fusion energy systems, and the engineering evaluation of space industry and semiconductor products.
Numerical simulation methods for predicting plasma motion are generally categorized according to the spatial and temporal scale.
For instance, simulations at microscopic scales utilize Particle-In-Cell (PIC) simulations and Vlasov-Maxwell (or Vlasov-Poisson) simulations, while at macroscopic scales, Magnetohydrodynamic (MHD) simulations, which involve static relaxation and fluid approximations, are employed.
As computing technology has advanced, PIC-MHD and Vlasov-MHD simulations that connect these hierarchical scales, as well as low-resolution Full-PIC and Full-Vlasov simulations, have been developed.
This has led to a better understanding of multi-scale plasma phenomena, such as magnetic reconnection.
However, none have achieved significant increases in simulation size due to the immense computational costs involved.
It is an important task to seek methods that can consistently simulate all multi-scale plasma physical phenomena.
This necessitates the development of computational methods capable of simultaneously handling microscopic kinetic effects and macroscopic MHD turbulence.

We focus on the Vlasov-Maxwell equations in plasma simulations.
The primary challenge of Vlasov simulations is the high computational cost required for the time evolution of the six-dimensional partial differential equations.
Assuming that the Vlasov equation can be divided into six advection equations and that each advection equation is solved using a simple explicit method, the computational complexity of classical calculations becomes $O(N^6(T^2/\epsilon+cT/\Delta x))$~\citep{Sato2024}.
Here, $N$ is the number of isotropic and uniform grids for each six-dimensional direction, $T$ is the simulation time, $\epsilon$ is the error tolerance, $c$ is the advection velocity, and $\Delta x$ is the spatial grid width.

In recent years, the development of quantum algorithms for classical partial differential equations (PDEs) assuming fault-tolerant quantum computers has rapidly progressed.
Notable examples include algorithms for the Navier-Stokes equations~\citep{Gaitan2020,Gaitan2021}, the Vlasov equation~\citep{Engel2019,Ameri2023,Toyoizumi2024,Miyamoto2024}, the Poisson equation~\citep{Cao2013,Wang2020}, and the wave equation~\citep{Costa2019,Suau2021}.
Quantum algorithms for linear ordinary differential equations, such as the HHL algorithm~\citep{Harrow2009,Childs2017} and Hamiltonian simulation algorithms~\citep{Berry2007,Childs2010,Berry2012,Childs2012,Berry2014,Berry2015b,Berry2015a,Berry2016,Novo2017,Childs2018}, have been proposed to potentially provide exponential speedup.
Efficient implementation of the Hamiltonian simulation has mainly utilized the Trotter decomposition algorithm~\citep{Jin2023a,Jin2023b}, or Quantum Singular Value Transformation~(QSVT) algorithm~\citep{Gilyen2019}.
\citep{Sato2024} demonstrated quantum computation of advection and wave equations using Hamiltonian simulation with Trotter decomposition, highlighting issues with spatial discretization using central difference and errors unique to Trotter decomposition from a numerical computation scheme perspective.
\citep{Hu2024} illustrated a method for “Schr\"{o}dingerization” of spatial discretization using upwind differences under specific conditions.
There is a growing trend towards enhancing the accuracy of quantum computational methods in line with the history of classical numerical computation.

Various plasma physics problems have been explored for potential applications to quantum computing~\citep{Dodin2021}.
Plasma simulations are typically formulated as coupled sets of nonlinear partial differential equations, and thus transforming them into linear Schr\"{o}dinger equations is not straightforward.
However, wave problems, such as three-wave interactions in plasma~\citep{Shi2021} and extraordinary waves in cold plasma (electromagnetic waves with electric fields perpendicular to the background magnetic field)~\citep{Novikau2022}, have been efficiently mapped to Schr\"{o}dinger equations.
Ideally, a quantum algorithm capable of solving the first-principles description of plasma physics provided by the Vlasov equation is desired.
For problems with weak nonlinearity, implementation methods for linearizing and performing Hamiltonian simulation on the Vlasov-Maxwell equations~\citep{Engel2019}, the collisional Vlasov-Poisson equations~\citep{Ameri2023}, and the Vlasov-Poisson equations~\citep{Toyoizumi2024} have been demonstrated within the framework of spectral methods under static assumptions.
Specifically,~\citep{Toyoizumi2024} employs the Hamiltonian simulation based on QSVT, allowing the application to different problems by merely changing the block-encoding quantum circuits of the Hamiltonian.

However, linearizing the nonlinear Vlasov-Maxwell equations using quantum algorithms for nonlinear differential equations is challenging.
For example, both Carleman Linearization~\citep{Carleman1932,Kowalski1991,Liu2021} and Koopman-von Neumann Linearization~\citep{Koopman1931,Neumann1932} are hindered by the first moment calculation term of the current density within the system of equations.
Consequently, among the existing quantum algorithms, solving the genuinely nonlinear Vlasov-Maxwell equations requires either a yet-to-be-proposed quantum algorithm specialized for the first moment calculation of nonlinear differential equations or a sequential approach where the Vlasov and Maxwell equations are time-evolved separately, with the results measured and combined on a classical computer.

This study implements a quantum-classical hybrid algorithm where the QSVT-based Hamiltonian simulation of the Vlasov equation and the classical computation of the Maxwell equations are executed sequentially.
At each time step, the solution from the Vlasov solver is extracted from the quantum node and interacted with the Maxwell solver on the classical node.
It should be noted that this method requires considering arbitrary state preparation at each step and quantum amplitude estimation for the number of grid points in the three-dimensional real space, so strict quantum advantage is not claimed.
This study aims to evaluate the impact on numerical solutions and stability conditions when using the QSVT-based Hamiltonian simulation as a subroutine of the Vlasov solver.
Chapter 2 introduces the numerical schemes embedded in quantum computation and the implementation methods of QSVT.
In Chapter 3, we present specific block-encoding quantum circuits for the Hamiltonian required for the QSVT-based Hamiltonian simulation of the Vlasov-Maxwell equations tailored to specific problem settings.
Chapter 4 details the construction of quantum circuits on a classical emulator using Qiskit-Aer-GPU and the execution of quantum computations on an A100 GPU for 1D advection tests and 1D1V two-stream instability conditions. To compare the results of quantum computations, classical matrix exact diagonalization, and Trotter decomposition-based Hamiltonian simulations were also performed.
This comparison clarifies the differences in scheme errors between QSVT and Trotter decomposition, and the accuracy of computations compared to classical methods.
The effectiveness of QSVT-based Hamiltonian simulation as a numerical computation method in quantum schemes is quantitatively evaluated.
In Chapter 5, we discuss the characteristics of the numerical schemes embedded in quantum computation and the numerical stability conditions arising from scheme errors in QSVT-based Hamiltonian simulation, from the perspective of numerical computation schemes.

\section{PRELIMINARY}
\subsection{Semi-discretized central difference scheme\label{sec:semi_centra_scheme}}
Let us consider a six-dimensional phase space $\Omega := [0, L] \times [0, L] \times [0, L] \times [-V, V] \times [-V, V] \times [-V, V]$.
Here, $L, V \in \mathbb{R}_+$, and for simplicity, we assume that the three-dimensional physical space is uniform in the range $[0, L]$ and the three-dimensional velocity space is uniform in the range $[-V, V]$.
Each direction of the domain is discretized with $N$ uniformly distributed points.
Thus the grid interval are $\Delta x = \Delta y = \Delta z = L / N$ and $\Delta v := \Delta v_x = \Delta v_y = \Delta v_z = 2V / N$ where $N = 2^n$ and $n \in \mathbb{N}$.

We define the six-dimensional scalar $f(\boldsymbol{x},\boldsymbol{v})$ in the phase space $\Omega$.
We hereafter label the grid points in $\Omega$ with vectors of indices:
\begin{equation}
    \boldsymbol{j}:=(j_x,j_y,j_z,j_{v_x},j_{v_y},j_{v_z})\in \mathcal{I}_6:=[N]^{\otimes 6},
\end{equation}
where, $j_x, j_y,$ and $j_z$ represent the index of the grid points in the three-dimensional physical space, and $j_{v_x}, j_{v_y},$ and $j_{v_z}$ represent the indices of the grid points in the three-dimensional velocity space, with each indices taking value from $0$ to $N-1$. 
For convenience in later use, we define a flattened index for the multi-dimensional information as 
\begin{equation}
    j:=j_x+j_yN+j_zN^2+j_{v_x}N^3+j_{v_y}N^4+j_{v_z}N^5.
\end{equation}

Now, the Vlasov equation for collisionless plasma is represented in a conservative form with the plasma distribution function in phase space as $f(\boldsymbol{x},\boldsymbol{v},t)$:
\begin{equation}
    \frac{\partial f(\boldsymbol{x},\boldsymbol{v},t)}{\partial t} + \boldsymbol{v}\cdot\frac{\partial f(\boldsymbol{x},\boldsymbol{v},t)}{\partial \boldsymbol{x}} + \frac{q}{m}(\boldsymbol{E}(\boldsymbol{x},t)+\boldsymbol{v}\times\boldsymbol{B}(\boldsymbol{x},t))\cdot\frac{\partial f(\boldsymbol{x},\boldsymbol{v},t)}{\partial \boldsymbol{v}} = 0,\label{eq:Vlasov-Maxwell}
\end{equation}
Here, $\boldsymbol{E}$,$\boldsymbol{B}$ and $\boldsymbol{v}$ are three-dimensional vectors, representing the electric field, magnetic field, and velocity in phase space, respectively.
To discretize the spatial derivative terms, we use the central difference method.
For example, the spatial derivative along the $x$ direction can be written with second-order accuracy as 
\begin{equation}
    \frac{\partial f(\boldsymbol{x},\boldsymbol{v})}{\partial x} = \frac{1}{2\Delta x}\left(f(\boldsymbol{x}+\Delta x\boldsymbol{e},\boldsymbol{v})-f(\boldsymbol{x}-\Delta x\boldsymbol{e},\boldsymbol{v})\right)+O\left(\Delta x^2\right),
\end{equation}
$\boldsymbol{e}:=(1,0,0)$ is the unit vector in physical space, and similarly for the derivatives with respect to $y$, $z$, $v_x$, $v_y$, and $v_z$. The Vlasov equation discretized at spatial grid points using second-order accurate central differences can be expressed as
\begin{align}
    \frac{\partial f(\boldsymbol{x},\boldsymbol{v},t)}{\partial t}\approx
    &- \frac{v_x}{2\Delta x}\left(f(x+\Delta x,y,z,\boldsymbol{v},t)-f(x-\Delta x,y,z,\boldsymbol{v},t)\right)\no\\
    & - \frac{v_y}{2\Delta x}\left(f(x,y+\Delta x,z,\boldsymbol{v},t)-f(x,y-\Delta x,z,\boldsymbol{v},t)\right)\no\\
    & - \frac{v_z}{2\Delta x}\left(f(x,y,z+\Delta x,\boldsymbol{v},t)-f(x,y,z-\Delta x,\boldsymbol{v},t)\right)\no\\
    & - \frac{q}{m}\frac{(E_x(\boldsymbol{x},t)+(\boldsymbol{v}\times\boldsymbol{B})_x(\boldsymbol{x},t))}{2\Delta v}
    \left( \begin{array}{l}
    f(\boldsymbol{x},v_x+\Delta v,v_y,v_z,t)\\
    -f(\boldsymbol{x},v_x-\Delta v,v_y,v_z,t)
    \end{array}
    \right)\no\\
    & - \frac{q}{m}\frac{(E_y(\boldsymbol{x},t)+(\boldsymbol{v}\times\boldsymbol{B})_y(\boldsymbol{x},t))}{2\Delta v}
    \left( \begin{array}{l}
    f(\boldsymbol{x},v_x,v_y+\Delta v,v_z,t)\\
    -f(\boldsymbol{x},v_x,v_y-\Delta v,v_z,t)
    \end{array}
    \right)\no\\
    & - \frac{q}{m}\frac{(E_z(\boldsymbol{x},t)+(\boldsymbol{v}\times\boldsymbol{B})_z(\boldsymbol{x},t))}{2\Delta v}
    \left( \begin{array}{l}
    f(\boldsymbol{x},v_x,v_y,v_z+\Delta v,t)\\
    -f(\boldsymbol{x},v_x,v_y,v_z-\Delta v,t)
    \end{array}
    \right).\label{eq:disc_Vlasov-Maxwell}
\end{align} 

Introducing the vectorized value $\boldsymbol{f}(t)$ that represents the value of $f$ at each grid point, this equation can be transformed as follows:
\begin{align}
    \frac{d \boldsymbol{f}(t)}{d t}= A(t) \boldsymbol{f}(t).\label{eq:A(t)f(t)}
\end{align} 
Here $\boldsymbol{f}(t)$ is a $N^6$ dimensional vector that is defined as
\begin{align}
    \boldsymbol{f}_{j}(t) = f(j_x\Delta x,j_y\Delta x,j_z\Delta x,j_{v_x}\Delta v,j_{v_y}\Delta v,j_{v_z}\Delta v,t).
\end{align} 
We call a numerical scheme that discretizes only space using central differences, without discretizing time, a semi-discrete central difference scheme.
When expressing it in terms of each spatial derivative term, $A(t)$ can be written as
\begin{align}
    A(t) = A_x+A_y+A_z+A_{v_x}(t)+A_{v_y}(t)+A_{v_z}(t).\label{eq:A(t)}
\end{align} 
$A_x$ is an $N^6 \times N^6$ square matrix, which can be expressed as the tensor product of each spatial matrix:
\begin{align}
    A_x = -\frac{1}{2\Delta x}D_{\mathrm{per},x}\otimes I_y\otimes I_z\otimes \mathrm{diag}(v_x)\otimes I_{v_y}\otimes I_{v_z},\label{eq:A_x} 
\end{align}
where $I_x$ is an $N \times N$ identity matrix and $\mathrm{diag}(v_x)$ is an $N \times N$ diagonal matrix with phase velocities as its diagonal elements.
$D_{\mathrm{per},x}$ is an $N \times N$ central difference operator with the periodic boundary condition, which is defined as
\begin{align}
    D_{\mathrm{per},x} = \begin{pmatrix}
    0 & 1 & 0 & \cdots & 0 & 0 & -1 \\
    -1 & 0 & 1 & \cdots & 0 & 0 & 0 \\
    0 & -1 & 0 & \cdots & 0 & 0 & 0 \\
    \vdots & \vdots & \vdots & \ddots & \vdots & \vdots & \vdots \\
    0 & 0 & 0 & \cdots & 0 & 1 & 0 \\
    0 & 0 & 0 & \cdots & -1 & 0 & 1 \\
    1 & 0 & 0 & \cdots & 0 & -1 & 0
    \end{pmatrix}. \label{eq:D_per}
\end{align} 
From Eq.~(\ref{eq:D_per}) and Eq.~(\ref{eq:A_x}), it is clear that $A_x$ is an antisymmetric matrix. 
$A_{y}$ and $A_{z}$ are defined in the same manner and thus they are time-independent matrices. 

On the other hand, $A_{v_x}$ is written as
\begin{align}
    A_{v_x}(t) = &-\frac{1}{2\Delta v_x}\hat{E_x}(t)\otimes D_{\mathrm{per},v_x}\otimes I_{v_y}\otimes I_{v_z}\no\\
    &-\frac{1}{2\Delta v_x}\hat{B_z}(t)\otimes D_{\mathrm{per},v_x}\otimes \mathrm{diag}(v_y)\otimes I_{v_z}\no\\
    &+\frac{1}{2\Delta v_x}\hat{B_y}(t)\otimes D_{\mathrm{per},v_x} \otimes I_{v_y}\otimes\mathrm{diag}(v_z). \label{eq:A_v_x}
\end{align}
Here $\hat{E_x}(t), \hat{B_y}(t)$ and $\hat{B_z}(t)$ are $N^3 \times N^3$ diagonal matrices with the electric and magnetic fields in the three-dimensional physical space as its diagonal elements. 
The differential operator in velocity space, $A_{v_x}, A_{v_y}$, and $A_{v_z}$ are time-dependent.
In this paper, however, we assume they are time-independent over a small time interval $[t, t + \Delta t]$ and define the effective Hamiltonian at each time step.
That is, we use the time-independent Hamiltonian at the $n$-th time step as follows:
\begin{align}
    \hat{H}_{t_n} := i A(t=t_n), \label{eq:H}
\end{align}
where $t_n = n  \Delta t $ and $n \in \{ 0, T/\Delta t  - 1\}$, with the simulation time $T$.
Because the matrix $A$ is antisymmetry, this effective Hamiltonian is Hermitian. 

Next, we define the effective quantum state as the sum of the normalized distribution function $\boldsymbol{f}_j$ encoded as the amplitude of the computational basis $|j\rangle$ corresponding to each grid point $j$, normalized by $\|\boldsymbol{f}(t)\|$.
\begin{align}
    |\psi(t)\rangle = \sum_{\boldsymbol{j}\in \mathcal{I}_6}\frac{\boldsymbol{f}_j(t)}{\|\boldsymbol{f}(t)\|}|j\rangle.
\end{align}

In conclusion, we map the Vlasov equation to the Schr\"{o}dinger equation with the time-dependent Hamiltonian:
\begin{align}
    \frac{d }{d t}|\psi(t)\rangle &= -i\hat{H}(t) |\psi(t)\rangle,
\end{align}
and descretizing time, assuming that the Hamiltonian is constant over each time step, Eq.~(\ref{eq:H}).
At each time step, the solution for the Schr\"{o}dinger equation is given as 
\begin{align}
    |\psi(t_{n+1})\rangle = \exp(-i\hat{H}_{t_n}\Delta t)|\psi(t_n)\rangle.
\end{align}
Therefore, performing the time evolution iteratively, we finally obtain the general solution at any arbitrary time as
\begin{align}
    |\psi(t)\rangle = \prod_{n=0}^{N_t-1}\exp(-i\hat{H}_{t_n}\Delta t)|\psi(0)\rangle.
\end{align}
Here $N_t$ is the number of time steps, and the step indices are defined as $n \in \{0, 1, 2, \ldots, N_t-1\}$. This general solution is implemented using a quantum circuit.

\subsection{Quantum singular value transformation}
Quantum Singular Value Transformation (QSVT) is a quantum algorithm that applies polynomial transformations to the singular values of a given matrix ~\citep{Gilyen2019,Martyn2021}.
QSVT is versatile and can be applied to various quantum algorithms, including Hamiltonian simulation.
Specifically, it transforms the target function $\exp(-i\hat{H}\Delta t)$ into an approximating polynomial $P_R$ of degree $2R+1$, which is then implemented within a quantum circuit within the error tolerance $\epsilon$.

In this study, we follow the discussions of ~\citep{Toyoizumi2024} to formulate the necessary QSVT-based Hamiltonian simulation.
First, block encoding is a method that encodes a matrix $\hat{H}$ as the upper-left block of a unitary matrix $U$. For a normalization factor $\eta, \alpha$ and  the number of ancilla qubits $a$, $U$ is called an $(\eta\alpha, a, 0)$-block encoding of $\hat{H}$ if it satisfies the following condition:
\begin{align}
    U = \begin{pmatrix}
    \frac{\hat{H}}{\eta\alpha} & \cdot \\
    \cdot & \cdot
    \end{pmatrix}.
    \label{eq:U}
\end{align}
Here,$\eta$ is defined as the normalization factor resulting from the Hadamard gates applied to the ancilla qubits, which depends on the dimensional degree of each problem setup. For example, in the case of the 1D1V Vlasov equation,
And, the target matrix is embedded in the upper-left block of the unitary matrix, while the other blocks do not have physical significance.

When $\|\hat{H}\| \geq 1$, it is necessary to use the normalization factor to normalize the matrix for block encoding to maintain unitarity. 
As discussed in ~\citep{Toyoizumi2024}, this implementation introduces a time parameter $\tau$ and adjusts the time step width to $\Delta t=\tau/\eta\alpha$, thereby resolving any practical inconveniences in the implementation.

\begin{figure}
\begin{center}
\rotatebox{270}{
\includegraphics[scale=0.52]{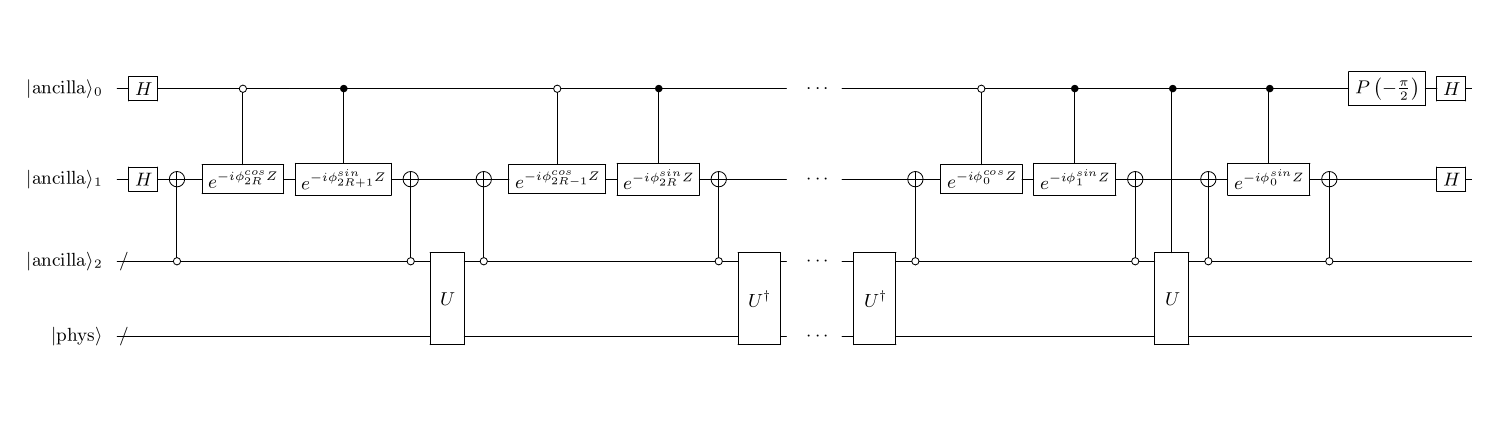}
}
\caption{The quantum circuit $U_{\mathrm{exp}}$block encodes the time evolution operator $\frac{1}{2}\exp(-i\hat{H}\Delta t)$ with $(1, a+2, \epsilon)$-block encoding. Provided by \citep{Toyoizumi2024}.}
\label{fig:Uexp}
\end{center}
\end{figure}

Next, using the implementation time parameter $\tau$ and the error $\epsilon$, $U_{\mathrm{exp}}$ satisfies the following condition as a $(1, a+2, \epsilon)$-block encoding of $\frac{1}{2}\exp\left(-i\frac{\hat{H}}{\eta\alpha}\tau\right)$.
\begin{align}
    \left\Vert\frac{\exp\left(-i\frac{\hat{H}}{\eta\alpha}\tau\right)}{2}-\langle \text{ancilla}|_{0,1,2}\otimes U_{\mathrm{exp}} \otimes \tanc_{0,1,2}\right\Vert\leq \epsilon.
    \label{eq:QSVT_expH}
\end{align}
Based on~\citep{Toyoizumi2024}, Fig.~\ref{fig:Uexp} shows the $(1, a+2, \epsilon)$-block encoding quantum circuit for $U_{\mathrm{exp}}$. The phase factor $\phi \in \mathbb{R}^{d+1}$ in the figure corresponds to the degree $d$ of the approximation polynomial.
The reason for the coefficient $\frac{1}{2}$ in $\exp\left(-i\frac{\hat{H}}{\eta\alpha}\tau\right)$ is that, as shown in Fig.~\ref{fig:Uexp}, we consider the attenuation of the amplitude of the solution state due to the superposition state of the ancilla qubit.
According to the Euler's formula, we have
\begin{align}
    \exp\left(-ix\tau\right) = \cos\left(x\tau\right)-i\sin\left(x\tau \right),
\end{align}
and we perform the Jacobi-Anger expansion as
\begin{align}
 & \cos \left(x \tau\right)=J_{0} (\tau)+2\sum _{k=1}^{\infty } (-1)^{k} J_{2k} (\tau)T_{2k} \left(x\right),\no\\
 & \sin \left(x \tau\right)=2\sum _{k=0}^{\infty } (-1)^{k} J_{2k+1} (\tau)T_{2k+1} \left(x \right),
\end{align}
where $J_{m} (\tau)$ is the $m$-th Bessel function of the first kind and $T_{k} (x)$ is the $k$-th Chebyshev polynomial of the first kind.
We define the truncated series at an index $R$ by
\begin{align}
P_{R}^{\cos}(x) & :=J_{0} (\tau)+2\sum _{k=1}^{R} (-1)^{k} J_{2k} (\tau)T_{2k} (x),\no\\
P_{R}^{\sin}(x) & :=2\sum _{k=0}^{R} (-1)^{k} J_{2k+1} (\tau)T_{2k+1} (x).
\label{eq:Difine_Pcos_Psin}
\end{align}
According to Lemma 57 of ~\citep{Gilyen2019}, the approximation polynomials $P_{R}^{\cos}(x)$ and $P_{R}^{\sin}(x)$ have an upper bound on the error under the following conditions.
\begin{align}
\left| \cos (x\tau)-P_{R}^{\cos}(x)\right|  & \leq \frac{5}{4}\left(\frac{e|\tau|}{4( R+1)}\right)^{( 2R+2)} ,\no\\
\left| \sin (x\tau)-P_{R}^{\sin}(x)\right|  & \leq \frac{5}{4}\left(\frac{e|\tau|}{2( 2R+3)}\right)^{( 2R+3)} .
\end{align}
Finally, subsituting $\tau=\eta\alpha \Delta t$ and $x$ with $\frac{\hat{H}}{\eta\alpha}$, we get the $(1, a+2, \epsilon)$-block encoding condition for $\frac{1}{2}\exp\left(-i\hat{H}\Delta t\right)$.
\begin{align}
&\left\Vert \frac{\exp (-i\hat{H}\Delta t)}{2}-\frac{\left( P_{R}^{\cos}\left( \hat{H}\Delta t\right) -iP_{R}^{\sin}\left( \hat{H}\Delta t\right)\right)}{2}\right\Vert \nonumber\\
&\hspace{25mm}\leq \frac{5}{8}\left(\frac{e|\eta\alpha \Delta t|}{4( R+1)}\right)^{( 2R+2)} +\frac{5}{8}\left(\frac{e|\eta\alpha \Delta t |}{2( 2R+3)}\right)^{( 2R+3)} , 
\label{eq:epsilon_QSVT}
\end{align}

\begin{align}
\left| \cos\left( \frac{\hat{H}}{\eta\alpha }\tau\right) -P_{R}^{\cos}\left( \frac{\hat{H}}{\eta\alpha }\tau\right)\right| \leq \varepsilon ,\left| \sin\left( \frac{\hat{H}}{\eta\alpha }\tau\right) -P_{R}^{\sin}\left( \frac{\hat{H}}{\eta\alpha }\tau\right)\right| \leq \varepsilon ,
\end{align}
where $0< \varepsilon < \frac{1}{e}$ and
\begin{align}
R( \tau ,\varepsilon ) & =\left\lfloor \frac{1}{2} r\left(\frac{e\tau}{2} ,\frac{5}{4} \varepsilon \right)\right\rfloor ,\no\\
r( \tau ,\varepsilon ) & \leq \mathcal{O}( \tau+\log( 1/\varepsilon ))
\end{align}
Therefore, the number of queries for $U_{\mathrm{exp}}$ as $\frac{1}{2}\exp\left(-i\hat{H}\Delta t\right)$ is given by
\begin{align*}
R+R+1 & =2\left\lfloor \frac{1}{2} r\left(\frac{e\eta\alpha \Delta t}{2}  ,\frac{5}{4} \varepsilon \right)\right\rfloor +1\\
 & \leq \mathcal{O}\left(\eta\alpha \Delta t +\log( 1/\varepsilon )\right).\label{eq:Query_QSVT}
\end{align*}
For more details see ~\citep{Gilyen2019}. According to~\citep{Toyoizumi2024}, amplitude amplification is necessary to cancel the factor $1/2$ in Eq.~(\ref{eq:epsilon_QSVT}) and obtain $\exp (-i\hat{H}\Delta t)$.
They present two types of QSVT-based algorithms to implement this: the oblivious amplitude amplification (OAA)~\citep{Gilyen2019} and the fixed-point amplitude amplification (FPAA) ~\citep{Martyn2021}.
The total number of queries for $\exp (-i\hat{H}\Delta t)$, including amplitude amplification, is given by $3\left(2R+1\right)$ for OAA and $D\left(2R+1\right)$ for FPAA.
The degree $D$ was given asymptotically in~\citep{Gilyen2019,Martyn2021}.

\section{Implementation of the quantum circuit for block encoding the effective Hamiltonian}
In this section, we present specific examples of quantum circuit implementations of the effective Hamiltonian block encoding $U$ tailored to problem settings ranging from 1D to 3D3V. The key point is that the spatial displacement in the differencing process can be achieved by the unitary operations of the quantum walk (Coin operator and Shift operator) as described in~\citep{Childs2009}.
We modified the block encoding of the Forward-Time Centered-Space (FTCS) scheme's difference equation for the Vlasov-Maxwell system using quantum walk embedding~\citep{Higuchi2023} and adopted it for the implementation of the effective Hamiltonian Eq.~(\ref{eq:U}) in this work.

\subsection{Quantum circuit for generating difference terms}

The implementation of the quantum circuit is based on the coin operator and shift operator, which are the operational elements of the quantum walk. 
The coin operator provides transition probabilities, while the shift operator handles the vertex movement.
See~\citep{Douglas2009} for more details.

We define the rotation gate $R(\theta)$ along the $Y$ axis as
\begin{align}
    R(\theta) \equiv e^{-iY\arccos(\theta)} .
\end{align}
In order to encode the coefficients of the difference term in Eq.~(\ref{eq:disc_Vlasov-Maxwell}) to the amplitudes of corresponding qubits, we apply the rotation gate to the $|0\rangle$ state as 
\begin{align}
    R(\theta)|0\rangle = \theta|0\rangle+\sqrt{1-|\theta|^2}|1\rangle,
\end{align}
so that we obtain the desired state $|0\rangle$ with the coefficient $\theta$ encoded as its amplitude.
In this encoding method, $\theta$ must be real and satisfy $|\theta| \leq 1$.
For general values including the cases with $|\theta| > 1$, we need to normalize it using the factor $\alpha$ defined in Eq.~(\ref{eq:U}). Specific choices of $\alpha$ will be represented in the following subsections.

Next, we perform the differentiation by applying $U_{\mathrm{Incr.}}$ and $U_{\mathrm{Decr.}}$ to the computational basis, which are defined as
\begin{eqnarray}
  U_{\mathrm{Incr.}} | j \rangle = | j+1 \rangle,\no\\
  U_{\mathrm{Decr.}} | j \rangle = | j-1 \rangle.\label{eq:u_incrdecr}
\end{eqnarray}
These are unitary operators when we suppose the periodic boundary condition:
\begin{eqnarray}
  U_{\mathrm{Incr.}} | N-1 \rangle = | 0 \rangle,\no\\
  U_{\mathrm{Decr.}} | 0 \rangle = | N-1 \rangle,\label{eq:periodic}
\end{eqnarray}
and can be implemented on a circuit as
\begin{align}
\raisebox{10mm}{
\raisebox{-9mm}{$ U_{\text{Incr.}}= $\quad}  
  \Qcircuit @C=1em @R=.7em {
  & \ctrl{1} &\ctrl{1}&\ctrl{1} & \gate{X}& \qw\\
  & \ctrl{1} & \ctrl{1} & \targ &\qw& \qw \\
 & \ctrl{1} & \targ & \qw & \qw & \qw\\
 & \targ & \qw & \qw & \qw & \qw
  }  \raisebox{-9mm}{\quad,\quad $ U_{\text{Decr.}}= $\quad}  
  \Qcircuit @C=1em @R=.7em {
  & \ctrlo{1} &\ctrlo{1}&\ctrlo{1} & \gate{X}& \qw\\
  & \ctrlo{1} & \ctrlo{1} & \targ &\qw& \qw \\
 & \ctrlo{1} & \targ & \qw & \qw & \qw\\
 & \targ & \qw & \qw & \qw & \qw 
  }}.
\label{eq:fig_shift}
\end{align}

As we correspond each computational basis with a grid point in our algorithm, incrementing or decrementing the computational basis corresponds to shifting the physical quantity at an arbitrary grid point by $\pm 1$.
In other words, with this operation, we obtain $f(\boldsymbol{x}\pm \boldsymbol{e}\Delta x)$.

\subsection{Quantum circuit implementation for block encoding the effective Hamiltonian}
\begin{figure}
\begin{center}
\raisebox{0pt}[15pt][75pt]{
\Qcircuit @C=0.5em @R=1.2em {
&\lstick{\tanc_2\hspace{1mm} }& \qw&\qw&\qw&\qw&\qw&\gate{R(\theta_v)}&\qw&\qw&\qw&\qw&\qw&\gate{R_z(3\pi)}&\qw\no\\
&\lstick{\tanc_3\hspace{1mm} }& \qw&\gate{H}&\qw&\gate{Z}&\qw&\qw&\qw&\ctrlo{1} &\qw&\ctrl{1}&\qw&\gate{H}&\qw\no\\
&\lstick{\tx\hspace{1mm}}& {/} \qw&\qw&\qw&\qw&\qw&\qw&\qw&\gate{U_\text{Incr.}}&\qw&\gate{U_\text{Decr.}}&\qw&\qw&\qw \no
}}
\caption{Quantum circuit for the block encoding of the effective Hamiltonian of the 1D Vlasov equation $\hat{H}_{1D}$.}
\label{fig:1DHamiltonian}
\end{center}
\end{figure}
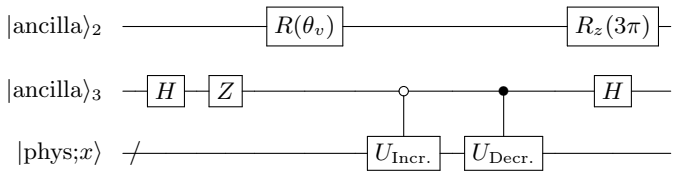

In this subsection we enumerate specific effective Hamiltonians to use and circuits for the implementation.
\begin{itemize}
\item Steady-state 1D Vlasov equation (advection equation):
\end{itemize}
\begin{equation}
    \frac{\partial f(x,t)}{\partial t} + v\frac{\partial f(x,t)}{\partial x} = 0.\label{eq:1D_Vlasov}
\end{equation}
When the phase velocity of the one-dimensional Vlasov equation is constant, it becomes equivalent to the steady-state advection equation. The advection equation describes the physics of the distribution function $f(x,t)$ propagating at the advection velocity $v$ while maintaining its waveform. By discretizing only the $x$ space using the central difference method to construct Eq.~(\ref{eq:A(t)f(t)}), we obtain $A = A_x$. Note that normalization is performed using the normalization coefficient $\alpha$ to satisfy the condition $|\theta_v| \leq 1$, where $\theta_v = v/2\Delta x \alpha = 1$. Referring to~\citep{Toyoizumi_Master}, based on Eq.~(\ref{eq:A_x}) and Eq.~(\ref{eq:H}), the effective Hamiltonian can be expressed as
\begin{align}
    \hat{H}_{1D} &= -i\theta_v D_{\mathrm{per},x},\no\\
                &= -i\theta_v \sum_{j_x=0}^{N-1}\left(|j_x\rangle \langle j_x+1|-|j_x\rangle \langle j_x-1|\right).\label{eq:H_1D}
\end{align}
From Fig.~\ref{fig:1DHamiltonian}, we can see that Hadamard gates are applied to the ancilla qubit, creating a superposition state.
This means that the sum in Eq.~(\ref{eq:H_1D}) is constructed on the ancilla qubit's $|0 \rangle_2$ state. Therefore, the amplitude is attenuated by a normalization factor $\eta=1/\sqrt{2}^2$, corresponding to the number of Hadamard gates.
Therefore, $(2\alpha,2,0)$-the block encoding of the effective Hamiltonian is implemented in the quantum circuit as shown in Fig.~\ref{fig:1DHamiltonian}.

\begin{itemize}
\item 1D1V Vlasov-Maxwell Equations without Magnetic Field:
\end{itemize}
\begin{equation}
    \frac{\partial f(x,v_x,t)}{\partial t} + v_x\frac{\partial f(x,v_x,t)}{\partial x} + \frac{q}{m}E(x,t)\frac{\partial f(x,v_x,t)}{\partial v_x} = 0,\label{eq:1D1V_Vlasov-Maxwell}
\end{equation}
\begin{equation}
 \frac{\partial E(x,t)}{\partial t}=-\frac{q}{m}\int vf(x,v,t) \, dv.\no
\end{equation}
The 1D1V Vlasov-Maxwell equations without a magnetic field describe the kinetic interaction between the plasma and the electric field. By discretizing only the phase space using the central difference method to construct Eq.~(\ref{eq:A(t)}), we obtain $A(t) = A_x + A_{v_x}(t)$. 

After discretizing the equation, we have the set of the coefficients $\{c_j\}$ with 
\begin{align}
    c_{j_{v_x}} &= \frac{v_{j_{v_x}}}{2\Delta x},\no\\
    c_{j_x} &= \frac{q}{m}\frac{E_x(x_{j_x},t)}{2\Delta v}.\label{eq:coef_1D1V}
\end{align}
Therefore, we take the normalization factor $\alpha$ to satisfy the condition $|\theta| \leq 1$ as
\begin{align}
    \alpha=\max\left|\frac{v}{2\Delta x}\right|+\max\left|\frac{q}{m}\frac{E_x(x,t)}{2\Delta v}\right|.\label{eq:a_1D1V}
\end{align}
Normalizing by this factor, we set $\theta$ as
\begin{align}
    \theta_{j_{v_x}} &= \frac{v_{j_{v_x}}}{2\Delta x \alpha},\no\\
    \theta_{j_x} &= \frac{q}{m}\frac{E_x(x_{j_x},t)}{2\Delta v\alpha}.\label{eq:theta_1D1V}
\end{align}

\begin{figure}
\begin{center}
\raisebox{0pt}[20pt][115pt]{
\Qcircuit @C=0.3em @R=1.2em {
&\lstick{\tanc_2\hspace{1mm} }& \qw&\qw&\qw&\qw&\qw&\qw&\qw&\gate{R(\theta_{j_x})}&\qw&\gate{R(\theta_{j_{v_x}})}&\qw&\qw &\qw&\qw&\qw&\qw&\qw&\qw&\qw&\gate{R_z(3\pi)}&\qw\no\\
&\lstick{\tanc_3\hspace{1mm} }& \qw&\gate{H}&\qw&\gate{Z}&\qw&\qw&\qw&\qw\qwx&\qw&\qw\qwx&\qw&\ctrlo{1} &\qw&\ctrl{1}&\qw&\ctrlo{1}&\qw&\ctrl{1}&\qw&\gate{H}&\qw\no\\
&\lstick{\tanc_4\hspace{1mm}}& \qw&\gate{H}&\qw&\qw&\qw&\qw&\qw&\ctrl{1}\qwx&\qw&\ctrlo{2}\qwx&\qw&\ctrlo{1} &\qw&\ctrlo{1}&\qw&\ctrl{2}&\qw&\ctrl{2}&\qw&\gate{H}&\qw\no\\
&\lstick{\tx\hspace{1mm}}& {/} \qw&\qw&\qw&\qw&\qw&\qw&\qw\qw&\gate{|j_x\rangle}&\qw&\qw&\qw&\gate{U_{\text{Incr.}}}&\qw&\gate{U_{\text{Decr.}}}&\qw&\qw&\qw&\qw&\qw&\qw&\qw \no\\
&\lstick{\tvx\hspace{1mm}}& {/} \qw&\qw&\qw&\qw&\qw&\qw&\qw&\qw&\qw&\gate{|j_{v_x}\rangle}&\qw&\qw&\qw&\qw&\qw&\gate{U_{\text{Incr.}}} &\qw&\gate{U_{\text{Decr.}}}\qw&\qw&\qw &\qw\no
}}
\caption{Quantum circuit for the block encoding of the effective Hamiltonian of the 1D1V Vlasov-Maxwell equations without magnetic field $\hat{H}_{1D1V}$.}
\label{fig:1D1VHamiltonian}
\end{center}
\end{figure}
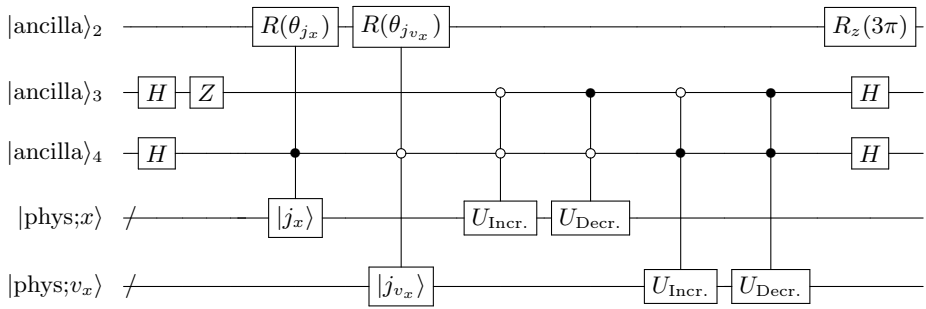

Based on Eq.~(\ref{eq:A_x}), Eq.~(\ref{eq:A_v_x}), and Eq.~(\ref{eq:H}), the effective Hamiltonian can be described using $\theta$ as follows:
\begin{align}
    \hat{H}_{1D1V} &= -\sum_{j_x=0}^{N-1}\sum_{j_{v_x}=0}^{N-1}
                \left(
                \begin{aligned}
                &i\theta_{j_{v_x}}(|j_x\rangle \langle j_x+1|-|j_x\rangle \langle j_x-1|)\otimes |j_{v_x}\rangle\langle j_{v_x}|\\
                &+i\theta_{j_x}|j_x\rangle \langle j_x|\otimes(|j_{v_x}\rangle \langle j_{v_x}+1|-|j_{v_x}\rangle\langle j_{v_x}-1|)
                \end{aligned}
                \right).
\end{align}
In a similar discussion as for the 1D Vlasov case, the amplitude is attenuated by a normalization factor of $\eta=1/\sqrt{2}^4$ due to the 4 Hadamard gates.
Therefore, the block encoding of the effective Hamiltonian for $(4\alpha,3,0)$ is implemented as shown in Fig.~\ref{fig:1D1VHamiltonian}.
The electric field is updated at each time step by solving Ampere's law from the Maxwell equations on a classical computer node.
The straightforward implementations of $\tphys$ states and the controlled $R(\theta_{j_x})$ and $R(\theta_{j_{v_x}})$ gates require $O(N)$ gates, which does not provide a quantum advantage.
According to~\citep{Toyoizumi2024}, the number of gates can potentially be improved to $O(\text{poly}(\log(N)))$ assuming Quantum Random Access Memory (QRAM)~\citep{Giovannetti2008}.

\begin{itemize}
\item 3D3V Vlasov-Maxwell Equations with Magnetic Field:
\end{itemize}
\begin{equation}
    \frac{\partial f(\boldsymbol{x},\boldsymbol{v},t)}{\partial t} + \boldsymbol{v}\cdot\frac{\partial f(\boldsymbol{x},\boldsymbol{v},t)}{\partial \boldsymbol{x}} + \frac{q}{m}(\boldsymbol{E}(\boldsymbol{x},t)+\boldsymbol{v}\times\boldsymbol{B}(\boldsymbol{x},t))\cdot\frac{\partial f(\boldsymbol{x},\boldsymbol{v},t)}{\partial \boldsymbol{v}} = 0,\label{eq:3D3V_Vlasov-Maxwell}
\end{equation}
\begin{align}
 \frac{\partial \boldsymbol{E}(\boldsymbol{x},t)}{\partial t} &= c^2\nabla\times\boldsymbol{B}(\boldsymbol{x},t)-\frac{q}{m}\int \boldsymbol{v}f(\boldsymbol{x},\boldsymbol{v},t)d\boldsymbol{v},\no\\
 \frac{\partial \boldsymbol{B}(\boldsymbol{x},t)}{\partial t} &= -\nabla\times\boldsymbol{E}(\boldsymbol{x},t).
\end{align}
The 3D3V Vlasov-Maxwell equations with a magnetic field fully reproduce the kinetic interaction between the plasma and the electromagnetic field. By discretizing only the 3D3V-phase space using the central difference method, we obtain Eq.~(\ref{eq:A(t)}). The normalization coefficient $\alpha$ is chosen to satisfy the condition $|\theta| \leq 1$ as
\begin{align}
    \alpha=\max\left|\frac{\boldsymbol{v}}{2\Delta x}\right|+\max\left|\frac{q}{m}\frac{\boldsymbol{E}(\boldsymbol{x},t)+\boldsymbol{v}\times \boldsymbol{B}(\boldsymbol{x},t)}{2\Delta v}\right|.\label{eq:a_3D3V}
\end{align}
This is defined as the sum of the maximum absolute values of the coefficients of each discretized term. And then, we set $\theta$ as
\begin{align}
    \theta_{j_{v_x}}&=\frac{v_{j_x}}{2\Delta x \alpha},\no\\
    \theta_{j_{v_y}}&=\frac{v_{j_y}}{2\Delta x \alpha},\no\\
    \theta_{j_{v_z}}&=\frac{v_{j_z}}{2\Delta x \alpha},\no\\
    \theta_{\boldsymbol{j_x},j_{v_y},j_{v_z}}&=\frac{q}{m}\frac{E_x(\boldsymbol{x}_{\boldsymbol{j_x}},t)+v_{y,j_{v_y}}B_z(\boldsymbol{x}_{\boldsymbol{j_x}},t)-v_{z,j_{v_z}}B_y(\boldsymbol{x}_{\boldsymbol{j_x}},t)}{2\Delta v\alpha},\no\\
    \theta_{\boldsymbol{j_x},j_{v_x},j_{v_z}}&=\frac{q}{m}\frac{E_y(\boldsymbol{x}_{\boldsymbol{j_x}},t)+v_{z,j_{v_z}}B_x(\boldsymbol{x}_{\boldsymbol{j_x}},t)-v_{x,j_{v_x}}B_z(\boldsymbol{x}_{\boldsymbol{j_x}},t)}{2\Delta v\alpha},\no\\
    \theta_{\boldsymbol{j_x},j_{v_x},j_{v_y}}&=\frac{q}{m}\frac{E_z(\boldsymbol{x}_{\boldsymbol{j_x}},t)+v_{x,j_{v_x}}B_y(\boldsymbol{x}_{\boldsymbol{j_x}},t)-v_{y,j_{v_y}}B_x(\boldsymbol{x}_{\boldsymbol{j_x}},t)}{2\Delta v\alpha}.\label{eq:theta_3D3V}
\end{align}
Here, the index $\boldsymbol{j_x}:=(j_x,j_y,j_z)$ is defined as $\in \mathcal{I}_3:=[N]^{\otimes 3}$. 
Therefore, based on Eq.~(\ref{eq:A_x}), Eq.~(\ref{eq:A_v_x}), and Eq.~(\ref{eq:H}),  the effective Hamiltonian can be expressed using $\theta$ as
\begin{align}
    \hat{H}_{3D3V} 
    = -\sum_{\boldsymbol{j}\in \mathcal{I}_6}^{N^6-1}
    \left( 
   \begin{aligned}
    &i\theta_{j_{v_x}}\left(\hat{T}_{j_x,j_y,j_z,j_{v_x},j_{v_y},j_{v_z}}^{(x,+1)}-\hat{T}_{j_x,j_y,j_z,j_{v_x},j_{v_y},j_{v_z}}^{(x,-1)}\right)\\
    +&i\theta_{j_{v_y}}\left(\hat{T}_{j_x,j_y,j_z,j_{v_x},j_{v_y},j_{v_z}}^{(y,+1)}-\hat{T}_{j_x,j_y,j_z,j_{v_x},j_{v_y},j_{v_z}}^{(y,-1)}\right)\\
    +&i\theta_{j_{v_z}}\left(\hat{T}_{j_x,j_y,j_z,j_{v_x},j_{v_y},j_{v_z}}^{(z,+1)}-\hat{T}_{j_x,j_y,j_z,j_{v_x},j_{v_y},j_{v_z}}^{(z,-1)}\right)\\    +&i\theta_{\boldsymbol{j_x},j_{v_y},j_{v_z}}\left(\hat{T}_{j_x,j_y,j_z,j_{v_x},j_{v_y},j_{v_z}}^{(v_x,+1)}-\hat{T}_{j_x,j_y,j_z,j_{v_x},j_{v_y},j_{v_z}}^{(v_x,-1)}\right)\\    +&i\theta_{\boldsymbol{j_x},j_{v_x},j_{v_z}}\left(\hat{T}_{j_x,j_y,j_z,j_{v_x},j_{v_y},j_{v_z}}^{(v_y,+1)}-\hat{T}_{j_x,j_y-1,j_z,j_{v_x},j_{v_y},j_{v_z}}^{(v_y,-1)}\right)\\  +&i\theta_{\boldsymbol{j_x},j_{v_x},j_{v_y}}\left(\hat{T}_{j_x,j_y,j_z,j_{v_x},j_{v_y},j_{v_z}}^{(v_z,+1)}-\hat{T}_{j_x,j_y-1,j_z,j_{v_x},j_{v_y},j_{v_z}}^{(v_z,-1)}\right)
    \end{aligned}
    \right),
\end{align}
where we define 
\begin{align}
    \hat{T}_{j_x,j_y,j_z,j_{v_x},j_{v_y},j_{v_z}}^{(x,\pm1)} := |j_x\rangle \langle j_x\pm1|\otimes |j_y\rangle\langle j_y|\otimes|j_z\rangle\langle j_z|\otimes|j_{v_x}\rangle\langle j_{v_x}|\otimes|j_{v_y}\rangle\langle j_{v_y}|\otimes|j_{v_z}\rangle\langle j_{v_z}|,
\end{align}
and similarly for the derivatives with respect to $y$, $z$, $v_x$, $v_y$ and $v_z$. 

\begin{figure}
\begin{center}
\rotatebox{90}{
\includegraphics[scale=0.75]{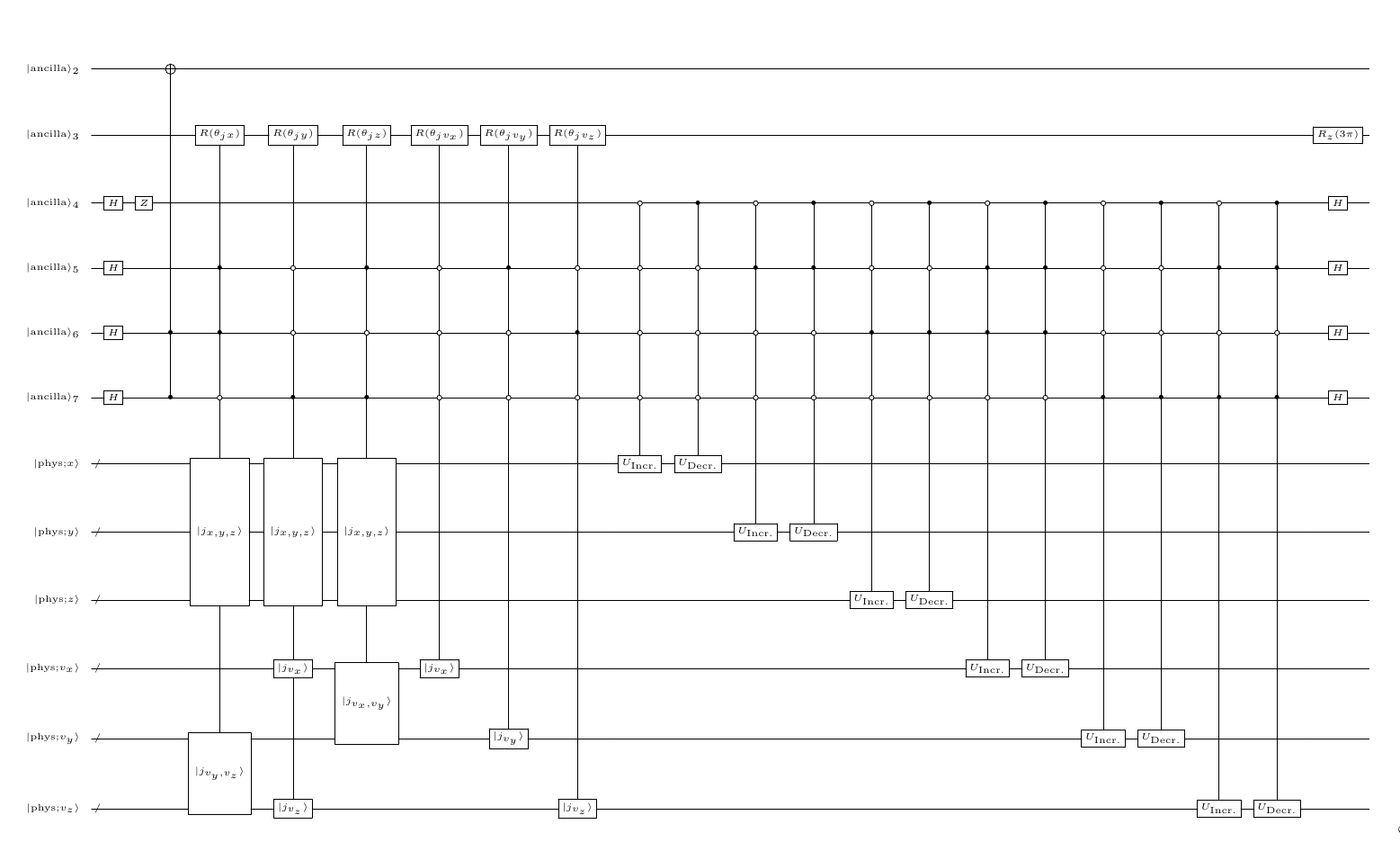}
}
\caption{Quantum circuit for the block encoding of the effective Hamiltonian of the 3D3V Vlasov-Maxwell equations with magnetic field $\hat{H}_{3D3V}$.}
\label{fig:3D3VHamiltonian}
\end{center}
\end{figure}

In a similar discussion as for the 1D Vlasov case, the amplitude is attenuated by a normalization factor of $\eta=1/\sqrt{2}^8$ due to the 8 Hadamard gates.
Therefore, the $(16\alpha,6,0)$-block encoding of the effective Hamiltonian is implemented as shown in Fig.~\ref{fig:3D3VHamiltonian}.
The electric and magnetic fields are updated at each time step by solving Ampere's law and Faraday's law from the Maxwell equations on a classical computer.
In the 3D3V case, we also assume that QRAM will be used to efficiently input the electromagnetic field information into the effective Hamiltonian at each time step.

\begin{figure}
\begin{center}
\includegraphics[scale=0.55]{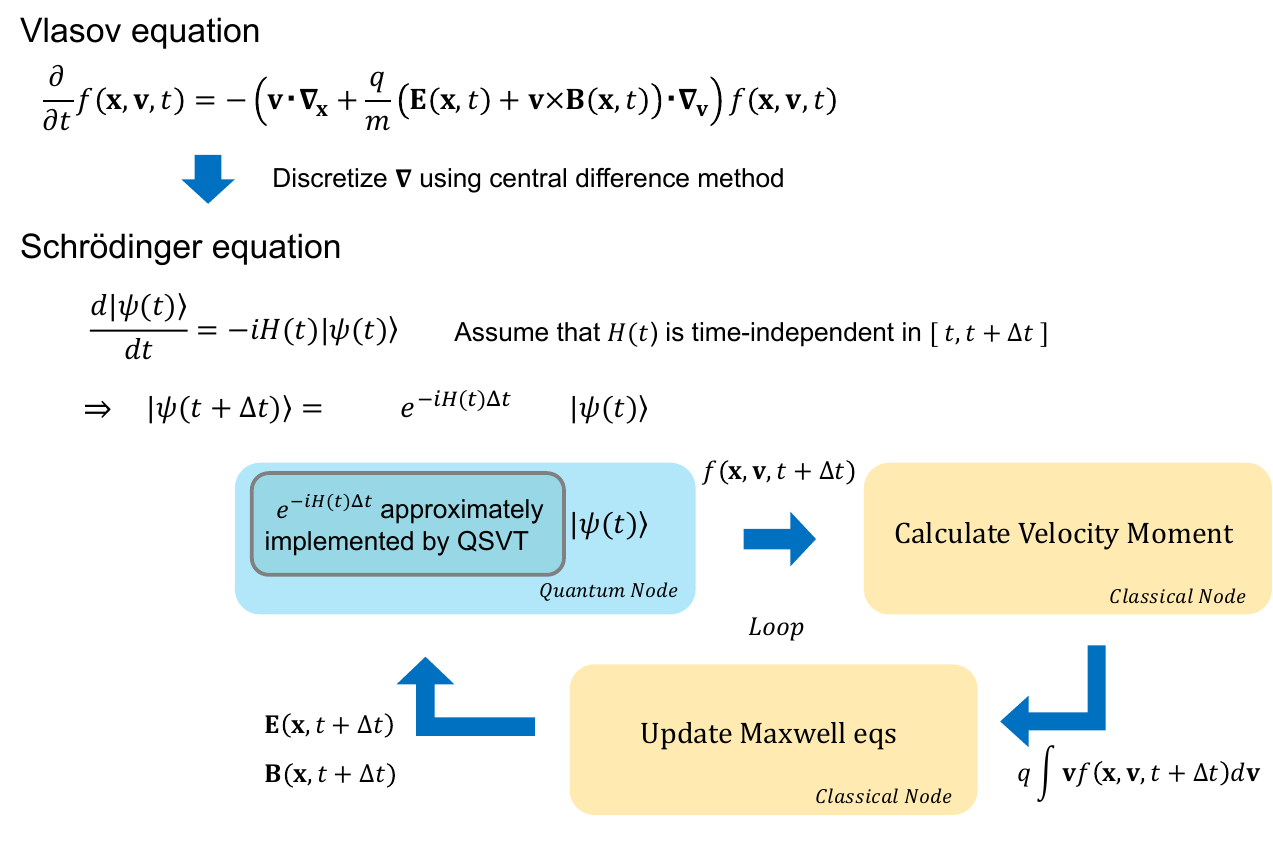}
\caption{Our numerical simulation algorithm flow. The Vlasov equation is simulated on a quantum circuit using Hamiltonian simulation based on QSVT, while the velocity moment calculation and Maxwell equations are executed classically. Additionally, for comparison, we simulated by replacing the exponential matrix with Trotter decomposition of the quantum scheme and exact diagonalization on the classical node.}
\label{fig:flow_method}
\end{center}
\end{figure}

\section{Numerical Results}
In this section, we provide numerical results for QSVT-based Hamiltonian simulation and several numerical schemes. We implemented the QSVT-based Hamiltonian simulation, referring to~\citep{Toyoizumi2024}. To evaluate the quantum scheme as a numerical computation scheme for the Vlasov equation, we prepared two problem settings: a 1D advection test and a 1D1V two-stream instability. We used the open-source \texttt{Qiskit 0.212}~\citep{Qiskit} as the quantum computing tool for implementing the quantum circuits. 
In this work, we used \texttt{pyqsp}~\citep{Chao2021} to calculate the phase factor $\phi$.
To save computational resources, we implemented the $(1,a+2,\epsilon)$-block encoding of $\frac{1}{2}\exp\left(-i\hat{H}\Delta t\right)$ as described in Eq.~(\ref{eq:epsilon_QSVT}) in a quantum circuit and extracted the complex probability amplitudes of arbitrary states using the \texttt{statevector\_simulator}. The obtained complex probability amplitudes were doubled and then multiplied by the normalization factor. Originally, as in~\citep{Toyoizumi2024}, a quantum amplitude amplification circuit should be applied to convert $\frac{1}{2}\exp$ to $\exp$ and extract the value, but considering the large scale of the quantum circuit for the two-stream instability case, this step was omitted.
\subsection{1D Advection Test}
To verify the properties of the QSVT-based Hamiltonian simulation as a numerical computation scheme, we conducted an advection test for the one-dimensional Vlasov equation with periodic boundary conditions. Additionally, for comparison, we implemented a first-order Trotter decomposition-based Hamiltonian simulation, referring to~\citep{Sato2024}, and performed the exact diagonalization.
\begin{itemize}
\item Simulation conditions of 1D Advection test:
\end{itemize}
\begin{align}
f(x, t=0) &= 1 \quad \text{for } 0 \leq x \leq 2^{n_x-1}, \notag \\
f(x, t=0) &= 0 \quad \text{otherwise}.
\end{align}
We set the number of $x$ qubits $n_x = 7$ (i.e., the number of grids $2^{n_x}=128$), the spatial interval $\Delta x = 1$, the time range $0 \leq t \leq 18$, the width of time step $\Delta t = 0.1$, and the advection velocity $v = 1$, resulting in the Courant number $\nu:=v\Delta t/\Delta x=0.1$. The initial distribution function represents a rectangular wave. Other conditions include the degree of the QSVT approximation polynomial: $R = 14$. The boundary condition is periodic.
\begin{figure}
\begin{center}
\includegraphics[scale=0.7]{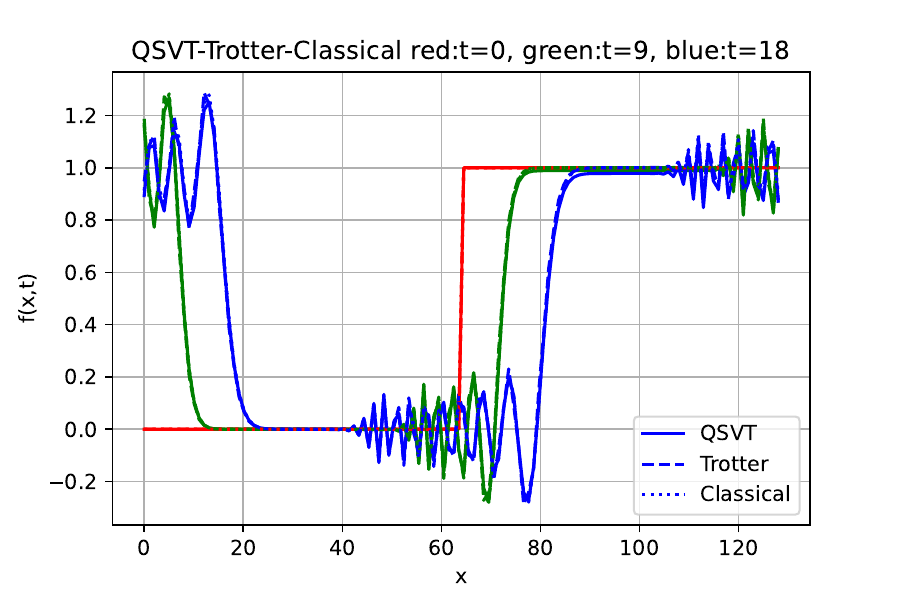}
\caption{The numerical results of each scheme for the 1D Vlasov equation under the advection test conditions are shown. The solid line represents QSVT, the dashed line represents Trotter decomposition, and the dotted line represents classical exact diagonalization. The red lines correspond to $t=0$, the green lines to $t=9$, and the blue lines to $t=18$, illustrating the time evolution.}
\label{fig:Adv_test}
\includegraphics[scale=0.7]{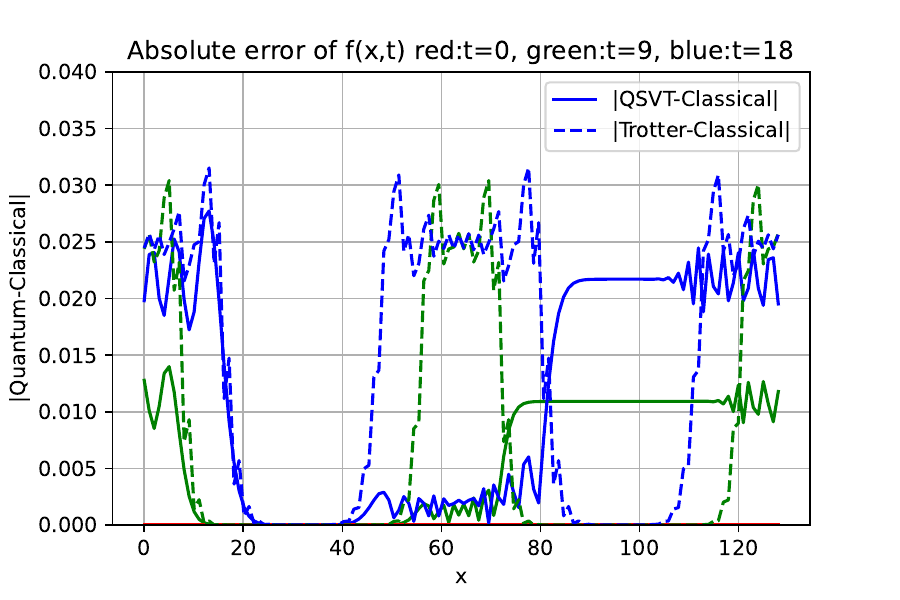}
\caption{The absolute errors between the numerical results of the quantum schemes and the classical exact diagonalization for the 1D Vlasov equation under the advection test conditions are shown. The solid line represents the absolute error between QSVT and Classical, and the dashed line represents the absolute error between Trotter decomposition and Classical.}
\label{fig:Adv_test_abusolute_error}
\end{center}
\end{figure}

Fig.~\ref{fig:Adv_test} shows the quantum computational results of the Hamiltonian simulations based on QSVT (solid line) and Trotter decomposition (dashed line), along with the numerical results of the classical exact diagonalization (dotted line).
The quantum circuits were executed on Qiskit-Aer simulator, and the real part of the amplitude of $|j\rangle$ embedded in $\boldsymbol{f}_j(t + \Delta t)$ was extracted using the \texttt{statevector\_simulator}. 
Fig.~\ref{fig:Adv_test} shows the advection propagation with $v = 1$ at $t = 0$ (red), $t = 9$ (green), and $t = 18$ (blue). 
Numerical oscillations induced by the discretization using the central difference method are observed in the nonlinear gradient regions. These numerical oscillations can be mitigated by using the upwind difference method or by adding numerical viscosity. 
This can potentially be implemented by transforming the effective Schr\"{o}dinger equation using the Schr\"{o}dingerization method proposed by~\citep{Hu2024}.

Next, Fig.~\ref{fig:Adv_test_abusolute_error} plots the absolute errors between QSVT and classical exact diagonalization (solid line), and between Trotter decomposition and the classical method (dashed line) (hereafter referred to as scheme error).
The error of QSVT is seen to be proportional to $|f(x_j, t)|$ and time $t$ at a given time $t$ at grid points $x_j$.
On the other hand, the error in Trotter decomposition is influenced by the high-frequency components of the numerical oscillations.
This indicates that significant differences in numerical results can arise depending on which quantum scheme is used.
We want to emphasize that, as we move towards practical applications, it is essential to select the appropriate quantum scheme based on the physical problem settings.
Moreover, since the results of Trotter decomposition and classical exact diagonalization are consistent with those presented in~\citep{Sato2024}, the validity of our code is assured.

\subsection{1D1V Two-Stream Instability Test}
Next, we focused on the two-stream instability test.
Historically, in the development of classical Vlasov solvers, the two-stream instability is one of the most fundamental nonlinear problems for the 1D1V Vlasov-Maxwell equations of plasma.
Whether this phenomenon can be accurately reproduced under appropriate conditions serves as a criterion for evaluating the numerical computation scheme~\citep{Filbet2003,Crouseilles2010,Qiu2011}.

We utilized a single node with an A100 GPU 40GB from the Computing Infrastructure Center at the University of Tsukuba, executing the quantum circuit simulator with \texttt{Qiskit-Aer-GPU}. The solutions to the Maxwell equations, computed on a classical node, were sequentially encoded into the effective Hamiltonian of the QSVT-based Hamiltonian simulation to solve the Vlasov-Maxwell system. 
Thus it should be noted that this algorithm is a quantum-classical hybrid algorithm.

\begin{itemize}
\item Simulation conditions of 1D1V two-stream instability based on~\citep{Qiu2011}:
\end{itemize}
\begin{align}
 f(x,v,t=0) &= \frac{2}{7\sqrt{2\pi}} \left(1 + \beta \frac{\cos(2kx) + \cos(3kx)}{1.2} + \cos(kx)\right) \exp\left(-\frac{v^{2}}{2}\right), \no\\
 E(x,t=0) &= 0.
\end{align}
We set the number of qubits for the $x$ and $v$ axis as $n_x = n_v = 5$, (i.e., simulation domain is defined the 1D1V phase space $32\times 32$), the time range $0 \leq t \leq 2500$, the time step $\Delta t = 1.89$, the $x$ space resolution $\Delta x = 0.39$, the $V$ space resolution $\Delta v = 0.31$, the initial perturbation amplitude parameter $\beta = 0.01$, and the initial wavenumber parameter $k = 0.5$.
The grid size was chosen to achieve approximately $ \nu \approx 1 $. The initial distribution function represents relative velocity in phase space. Other parameters were set as follows: the degree of the QSVT approximation polynomial $R = 9$, charge $q = 1$, mass $m = 1$, and permittivity $\epsilon_0 = 1$. The boundary condition for the $x$ physical space is periodic, while the boundary condition for the $v$ velocity space is fixed. The implementation method for fixed boundary conditions in QSVT is described in Appendix.~\ref{apendix:BC}. For simplicity, the current density is assumed to be the first moment of the distribution function for a single species of ions Eq.~(\ref{eq:1D1V_Vlasov-Maxwell}).\par
In this implementation, we directly extracted the time-evolved distribution function using QSVT as \texttt{statevector\_simulator} and calculated the current density by computing the moments classically. However, according to~\citep{Higuchi2023}, this method does not improve the computational cost of classical moment calculations and does not offer a quantum advantage. Therefore, instead of directly estimating the distribution function through Quantum Amplitude Estimation, we propose developing an estimation method that extracts the first-order moments using Quantum Numerical Integration.
\begin{figure}
\begin{center}
\includegraphics[scale=0.8]{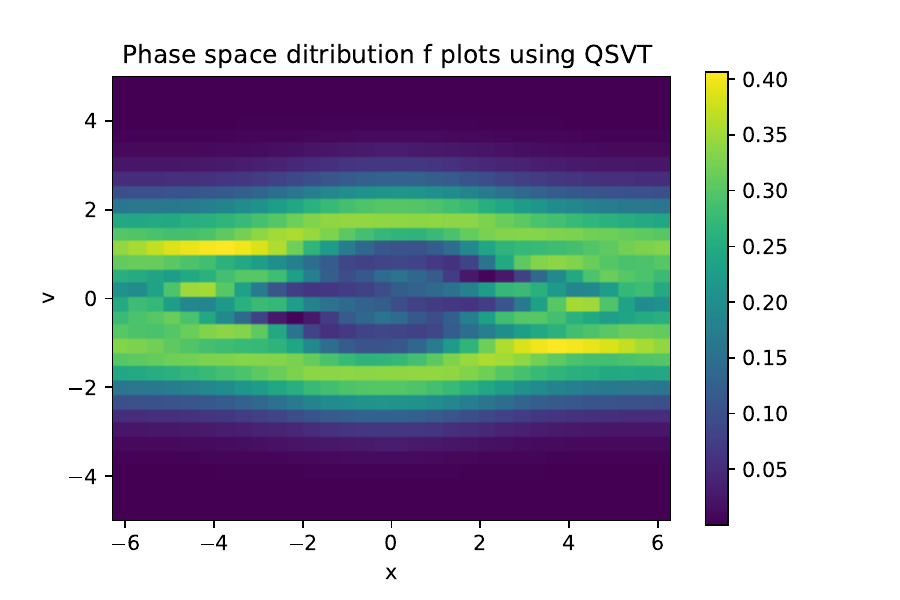}
\caption{The quantum computational numerical results of the QSVT-based Hamiltonian simulation scheme for the 1D1V Vlasov-Maxwell equations under the two-stream instability conditions. The distribution function in the $x$-$v$ phase space is shown at real time $T = 53$, with the vertical axis representing the velocity space and the horizontal axis representing the physical space.}
\label{fig:Twostream_test_QSVT}
\includegraphics[scale=0.8]{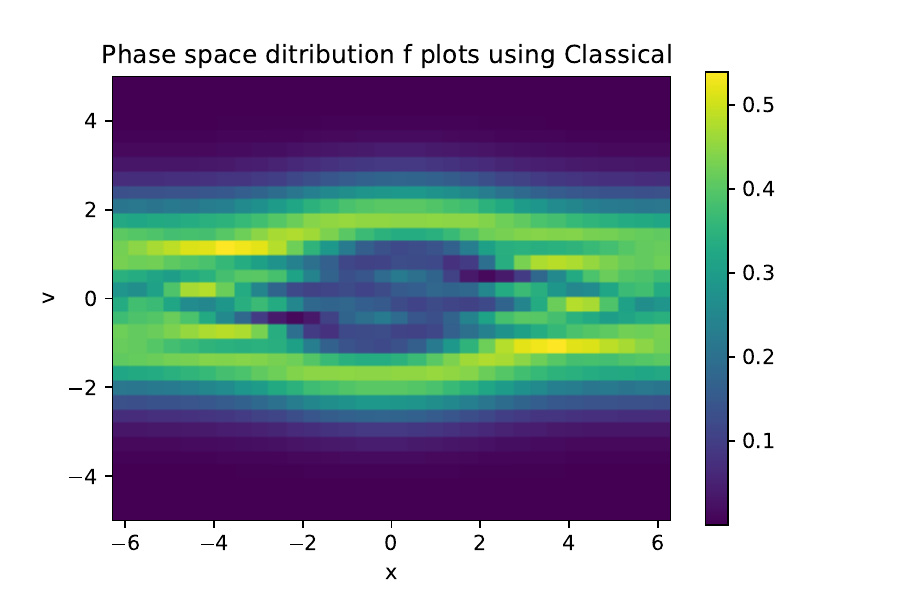}
\caption{The numerical results of the classical exact diagonalization scheme for the 1D1V Vlasov-Maxwell equations under the two-stream instability conditions. The distribution function in the $x$-$v$ phase space is shown at real time $T = 53$, with the vertical axis representing the velocity space and the horizontal axis representing the physical space.}
\label{fig:Twostream_test_Classical}
\end{center}
\end{figure}
Fig.~\ref{fig:Twostream_test_QSVT} shows the results of the two-stream instability at $T = 53$ for the 1D1V Vlasov-Maxwell equations using QSVT. The occurrence of phase mixing is evident. Fig.~\ref{fig:Twostream_test_Classical} shows the results obtained using classical exact diagonalization, and the vortex formation in the phase space is almost identical. However, focusing on the color bar, the values of the distribution function in QSVT are generally smaller than those in the classical method. This discrepancy arises from the quantum scheme error (approximation error) in QSVT's $ \exp(-i\hat{H}\Delta t) $ under the condition of Eq.~(\ref{eq:epsilon_QSVT}), leading to a difference from the true values. Nevertheless, as shown in Fig.~\ref{fig:Adv_test_abusolute_error}, the quantum scheme error in QSVT is proportional to the magnitude of the distribution function, thus not significantly affecting the vortex formation.
Therefore, the results between QSVT's Fig.~\ref{fig:Twostream_test_QSVT} and the classical method's Fig.~\ref{fig:Twostream_test_Classical} appear almost identical.

These results are also reasonably consistent with the vortex size and phase mixing depiction in the numerical results of the two-stream instability using the classical high-precision Vlasov scheme MPP SL DG method by~\citep{Qiu2011}. However, our results show the onset of numerical oscillations around the $x$ intervals $[-4,-2]$ and $[2,4]$. Over long-term evolution, these numerical oscillations are likely to induce different physical phenomena. As discussed in the 1D advection test, to ensure monotonicity, it is necessary to either introduce upwind differencing or incorporate numerical viscosity to suppress numerical oscillations and improve accuracy for nonlinear developments.

\section{Discussion}
Generally, the stability of numerical schemes for PDEs must be investigated. In particular, the properties of the semi-discretized central difference scheme embedded in the quantum system (Section.~\ref{sec:semi_centra_scheme}) are not well understood. Therefore, we discuss the stability of this scheme using von Neumann stability analysis. For simplicity, we take the 1D Vlasov equation as an example. When the distribution function $f(x,t)$ is expanded as a Fourier series in wavenumber space, it can be written as
\begin{align}
 f(x,t) &= \sum_k r_k(t)\exp(ikx),\\
 f_{j,k}^n &= f_k(x_j,t_n)=r^n\exp(ikx_j).
\end{align}
Here, $k$ is the wavenumber, $j$ is the physical space grid index, and $n$ is the time step index. Substituting it into Eq.~(\ref{eq:1D_Vlasov}), we obtain
\begin{align}
 \frac{\partial}{\partial t}r^n &= -i\frac{v}{\Delta x}\sin(k\Delta x)r^n.\label{eq:von_Nueman_1D}
\end{align}
However, assuming that $v$ is fixed within a small time interval $[t, t + \Delta t]$, as discussed in Section.~\ref{sec:semi_centra_scheme}, the amplitude $r$ is updated as
\begin{align}
 r^{n+1} &= \exp\left(-i\frac{v\Delta t}{\Delta x}\sin(k\Delta x)\right)r^n.
\end{align}
Thus, the amplification factor is expressed as
\begin{align}
 \left|\frac{r^{n+1}}{r^n}\right| = \left|\exp\left(-i\frac{v\Delta t}{\Delta x}\sin(k\Delta x)\right)\right|=1.\label{eq:r^n+1/r^n}
\end{align}
This always takes the value of 1 regardless of the classical Courant number($:=v\Delta t/\Delta x$).
In other words, the semi-discretized central difference scheme embedded in the quantum system is stable and free from numerical divergence.

Next, to investigate the phase shift (phase error) of propagation due to the semi-discretized central difference scheme, we perform a Fourier transform by setting $r_{\omega}^n = \exp(-i\omega t)$ in Eq.~(\ref{eq:von_Nueman_1D}). The resulting dispersion relation is as
\begin{align}
 \omega = \frac{v}{\Delta x}\sin(k\Delta x).\label{eq:dispersion}
\end{align}
Here, $\omega$ is the angular frequency. In the limit as $k\Delta x \rightarrow 0$, we have $\sin(k\Delta x) \approx k\Delta x$, which ensures convergence with the exact dispersion relation of the advection equation. Note that there is no constraint on the Courant number.

\begin{figure}
\begin{center}
\begin{tabular}{c@{\hspace{-4.5mm}}c@{\hspace{-4mm}}c}
\includegraphics[scale=0.315]{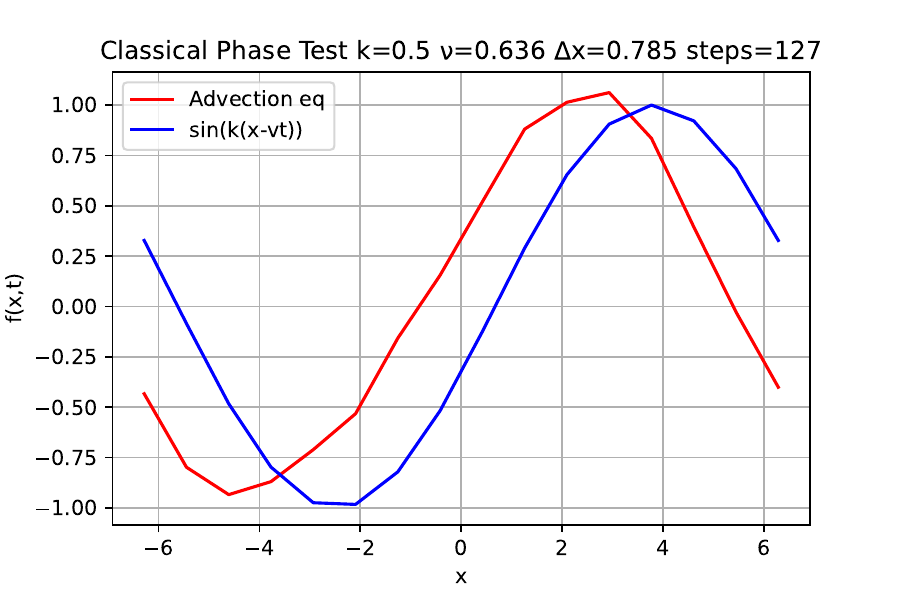}
&
\includegraphics[scale=0.315]{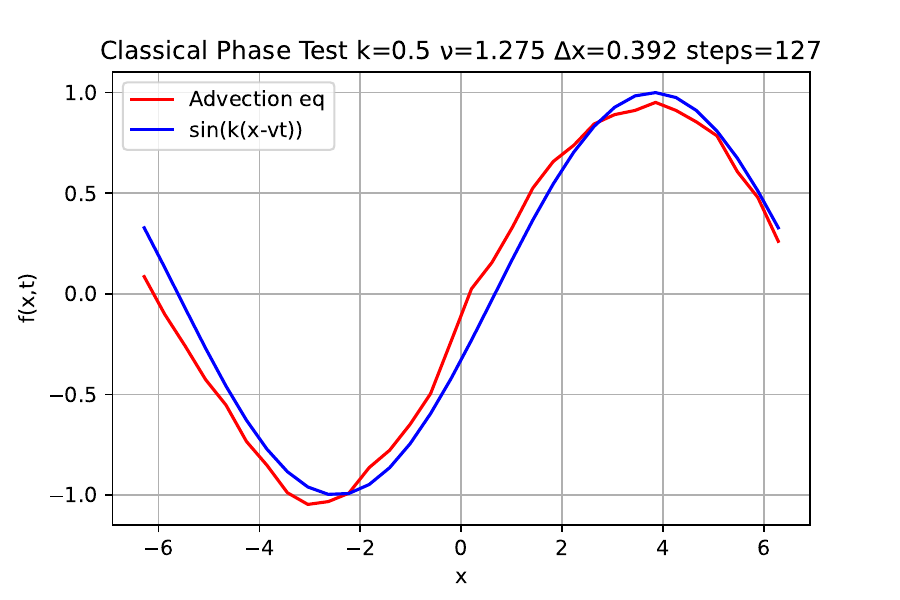}
&
\includegraphics[scale=0.315]{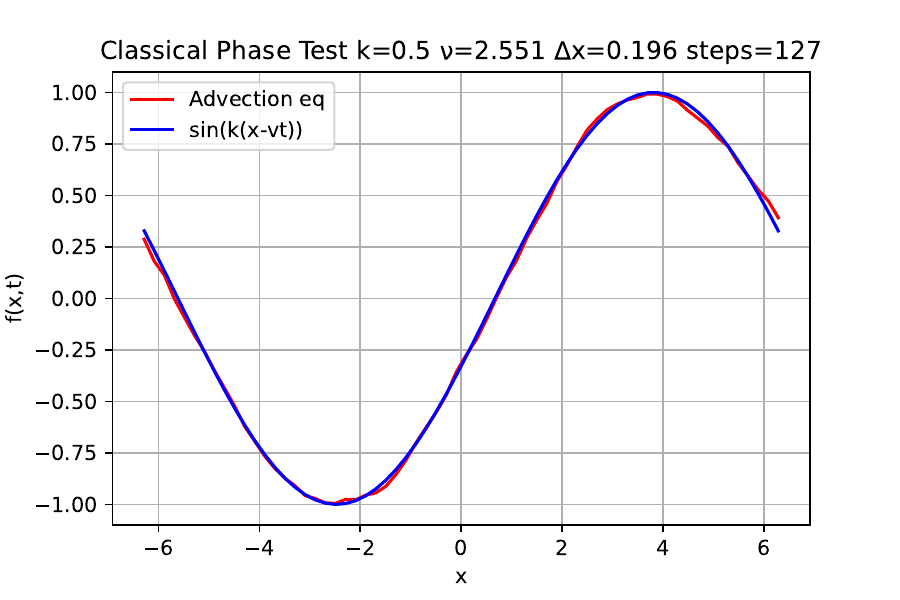}
\\
\hspace{2mm}(a)& 
\hspace{1mm} (b)&
\hspace{1mm} (c)
\end{tabular}
\caption{Numerical results of the semi-discretized central difference scheme (red line) for the 1D Vlasov equation and the exact solution (blue line) for the sine wave propagation test to investigate phase errors. (a): $\Delta x = 0.785$, $\nu = 0.636$; (b): $\Delta x = 0.392$, $\nu = 1.275$; (c): $\Delta x = 0.196$, $\nu = 2.551$; all displaying the results of propagation over the same real time under the same conditions.}
\label{fig:phase_error}
\end{center}
\end{figure}

In Fig. \ref{fig:phase_error}, to investigate the manifestation of phase errors, we compared the numerical solution of the 1D Vlasov equation for sinusoidal advection propagation with the exact solution of sinusoidal parallel propagation, while varying the Courant number along with $\Delta x$.
As anticipated from the dispersion relation Eq.~(\ref{eq:dispersion}), the figure shows that the phase error depends on $\Delta x$ rather than the Courant number $\nu$. 
Indeed, the numerical solution in Fig.~\ref{fig:phase_error}(c) showed the closest match to the exact solution under the conditions of the smallest $\Delta x$ and $\nu > 1$. This represents a significant advantage as a numerical scheme distinct from traditional classical numerical computations, such as finite difference methods.
In classical numerical schemes such as the finite difference method, the value of a grid point at the next time step is updated using several surrounding grid points (up to 7 points for high-accuracy methods) of an arbitrary point. 
This process is repeated for all grid points to obtain the information for the next time step across the entire grid. Such discrete updating methods must adhere to the CFL condition, which maintains the causality principle, meaning that the propagation speed of information cannot exceed the advection velocity. 

So, how does the semi-discretized central difference scheme embedded in the quantum system (Section.~\ref{sec:semi_centra_scheme}) determine which grid points at a given time are used to update the grid points at the next time step?
And what is the causality principle for this method? 
These questions are explained from the perspective of classical numerical schemes. 
In the semi-discretized central difference scheme embedded in the quantum system, the time evolution operator $\exp(-i\hat{H}\Delta t)$ acts on the state at the previous time step to update it to the next time step. 
For simplicity, consider a one-dimensional space divided into $N$ grid points, and let $f(x_j, t_n) = f_j^n $. By diagonalizing the time evolution operator $\exp(-i\hat{H}\Delta t)$ matrix for the semi-discretized central difference scheme of the advection equation, the information of the $j$-th grid point at step $n+1$ is described as
\begin{align}
 f_{j}^{n+1} = \omega_0(\nu)f_j^n-\sum_{r=1}^{N/2}\omega_r(\cos(\nu),\sin(\nu)) \left(f_{j+r}^n-f_{j-r}^n\right).
\end{align}
Here, the classical Courant number is $\nu = v \Delta t/\Delta x$. $\omega_r$ represents a weighting polynomial that includes $\cos(\nu)$ and $\sin(\nu)$ as variables, with the weights increasing the closer they are to $j$. The sum $\sum_{r=1}^{N/2}\omega_r=1$ is normalized. It becomes clear that the $\exp(-i\hat{H}\Delta t)$ matrix essentially has all its elements populated. This means that the semi-discretized central difference scheme embedded in the quantum system updates the information of any $j$-th grid point by referring to all grid point information from the previous time step, as illustrated in Fig.~\ref{fig:update_lattice}. We would like to consider a numerical interpretation of the properties of the weighting polynomial. Expanding $\exp(-i\hat{H}\Delta t)$ in a Taylor series around $\nu$, we get
\begin{align}
 \exp(-i\hat{H}\Delta t) &= \exp\left(\frac{\nu}{2}D_{\mathrm{per}}\right),\\
                        &= \sum_{s=0}^{\infty} \frac{\nu^s}{2^s s!}\left(D_{\mathrm{per}}\right)^s.
                        \label{eq:taylor_exp}
\end{align}
From Eq.~(\ref{eq:H}) and Eq.~(\ref{eq:A_x}), we use $\hat{H} = iA = ivD_{\mathrm{per}}/2\Delta x$.
When a Taylor approximation is made to $O(\nu^2)$, it matches the well-known classical finite difference method of the Forward-Time Centered-Space (FTCS) scheme. Moreover, for an accuracy of $O(\nu^{s+1})$ at the $s+1$th order, the update of the $j$th grid point uses interpolated values from within the range of $\pm s$ grid points around $j$. This has characteristics similar to implicit methods. Thus, the semi-discretized central difference scheme embedded in the quantum system refers to all grid point information from the previous time step and updates all grid points simultaneously, without violating causality even if it exceeds the CFL condition~\citep{Courant1928}. Quantum computation embeds this semi-discretized central difference scheme with a certain error tolerance. While classical computation tends to waste computational resources on diagonalization calculations, the scheme used for implementation in quantum computation provides a unique numerical advantage as a computational method.

In practice, however, it should be noted that we are not exactly implementing Eq.~(\ref{eq:taylor_exp}), but rather approximating it to a certain degree using quantum schemes such as QSVT or Trotter decomposition.
Consequently, the quantum schemes effectively reference grid points corresponding to the scaling order of the approximation error Eq.~(\ref{eq:epsilon_QSVT}).
For example, in the case of QSVT, the upper bound of the error is $O(2R)$, which means that the scheme references up to $\pm 2R$ around the $j$-th grid point.

\def\dx{1.5}
\begin{figure}
\begin{center}
\begin{tikzpicture}[scale=1.2]
    \node at (-1, 0) {$t+\Delta t$};
    \node at (-1, -1) {$t$};
    \draw (0, 0) -- (1*\dx, 0);
    \draw[dotted] (1*\dx, 0) -- (2*\dx, 0);
    \draw (2*\dx, 0) -- (4*\dx, 0);
    \draw[dotted] (4*\dx, 0) -- (5*\dx, 0);
    \draw (5*\dx, 0) -- (6*\dx, 0);
    \draw (0, -1) -- (1*\dx, -1);
    \draw[dotted] (1*\dx, -1) -- (2*\dx, -1);
    \draw (2*\dx, -1) -- (4*\dx, -1);
    \draw[dotted] (4*\dx, -1) -- (5*\dx, -1);
    \draw (5*\dx, -1) -- (6*\dx, -1);
    \node at (0*\dx, 0) {$\bullet$};
    \node at (1*\dx, 0) {$\bullet$};
    \node at (2*\dx, 0) {$\bullet$};
    \node at (3*\dx, 0) {$\bullet$};
    \node at (4*\dx, 0) {$\bullet$};
    \node at (5*\dx, 0) {$\bullet$};
    \node at (6*\dx, 0) {$\bullet$};
    \node at (3*\dx, 0.3) {$j$};
    \node at (0*\dx, -1) {$\bullet$};
    \node at (1*\dx, -1) {$\bullet$};
    \node at (2*\dx, -1) {$\bullet$};
    \node at (3*\dx, -1) {$\bullet$};
    \node at (4*\dx, -1) {$\bullet$};
    \node at (5*\dx, -1) {$\bullet$};
    \node at (6*\dx, -1) {$\bullet$};
    \node at (0*\dx, -1.3) {$0$};
    \node at (1*\dx, -1.3) {$1$};
    \node at (2*\dx, -1.3) {$j-1$};
    \node at (3*\dx, -1.3) {$j$};
    \node at (4*\dx, -1.3) {$j+1$};
    \node at (5*\dx, -1.3) {$N-2$};
    \node at (6*\dx, -1.3) {$N-1$};
    \draw[->, thin, >=stealth] (0*\dx, -1) -- (3*\dx, 0);
    \draw[->, thin, >=stealth] (1*\dx, -1) -- (3*\dx, 0);
    \draw[->, thin, >=stealth] (2*\dx, -1) -- (3*\dx, 0);
    \draw[->, thin, >=stealth] (3*\dx, -1) -- (3*\dx, 0);
    \draw[->, thin, >=stealth] (4*\dx, -1) -- (3*\dx, 0);
    \draw[->, thin, >=stealth] (5*\dx, -1) -- (3*\dx, 0);
    \draw[->, thin, >=stealth] (6*\dx, -1) -- (3*\dx, 0);
\end{tikzpicture}
\caption{An illustration of the method for updating grid point information. The horizontal axis represents space, and the vertical axis represents time. Arrows indicate that the grid point information at the start point is used to update the grid point at the end point.}
\label{fig:update_lattice}
\end{center}
\end{figure}
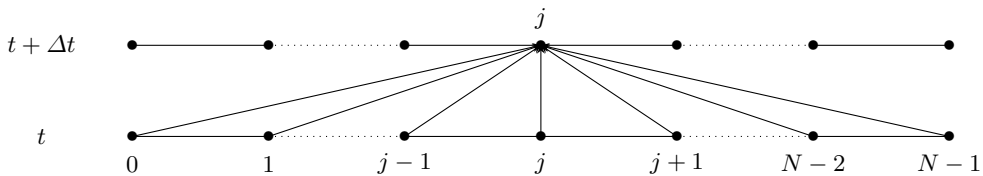

In classical numerical computations, the CFL condition is imposed as a stringent constraint on finite discretized numerical schemes~\citep{Courant1928}.
On the other hand, for the semi-discretized central difference scheme embedded in the quantum system, the CFL condition derived from von Neumann stability analysis is not imposed, as seen from Eq.~(\ref{eq:r^n+1/r^n}).
Does this mean that there is no numerical stability condition in quantum computation corresponding to the classical CFL condition (hereafter referred to as the "Quantum Stability Condition")?
The quantum stability condition can be expressed as follows using the error of implementation of $\exp(-iH\Delta t)$ based on QSVT Eq.~(\ref{eq:epsilon_QSVT}) and the error tolerance $\epsilon$:
\begin{align}
 \frac{5}{8}\left(\frac{e\eta\nu}{4(R+1)}\right)^{2R+2}+\frac{5}{8}\left(\frac{e\eta\nu}{2(2R+3)}\right)^{2R+3} \leq \epsilon.\label{eq:stability}
\end{align}
Here, the normalization coefficient for the 1D case is $\alpha=\nu/2$,$\eta=1/2$. For the 1D1V case, substitute Eq.~(\ref{eq:a_1D1V}), and for the 3D3V case, substitute Eq.~(\ref{eq:a_3D3V}). The quantum stability condition is broader concerning the Courant number compared to the classical condition. In the case of QSVT, the larger the degree $R$ in Eq.~(\ref{eq:stability}), the more relaxed the constraint on the Courant number becomes. However, as shown in Eq.~(\ref{eq:Query_QSVT}), increasing $R$ (i.e., reducing $\epsilon$) increases computational complexity. Therefore, relaxing the constraint on the Courant number is a trade-off with computational resources.  It is also possible to derive quantum stability conditions for other quantum schemes using similar procedures. For example, in the case of the Trotter decomposition used in the 1D advection test, the first-order Lie-Trotter-Suzuki decomposition-based approximation implementation error is $\left(\nu/2\right)^2 (n_x-1)/2$ as referenced in~\citep{Sato2024}. Additionally, for higher-order Trotter decompositions, the approximation implementation error depends on the order $p$ and is $O\left(\nu^{p+1}\right)$~\citep{Childs2021}.

The semi-discretized central difference scheme embedded in the quantum system in this work induces numerical oscillations in nonlinear regions due to Godunov's theorem.
To avoid the numerical oscillations, it is known to be useful to introduce an upwind difference scheme or numerical viscosity.
Quantum algorithms for PDEs that have attempted to introduce upwind difference schemes or viscosity terms (second-order derivatives) include~\citep{Hu2024}, but they have not achieved complete embedding.
In particular, for the upwind difference scheme, the issue lies in how to embed the mechanism to determine the flow direction.
If the upwind difference scheme can be embedded, first-order accuracy ensuring monotonicity is sufficient.
One of the significant advantages of quantum computing is that it can prepare an exponentially large number of grids relative to the number of qubits, allowing grid resolution to be made arbitrarily small under the quantum CFL condition.
Therefore, implementing upwind difference schemes and numerical viscosity in a quantum framework becomes a necessary condition for solving high-resolution shock wave problems and nonlinear instability development problems of turbulence in plasma simulations with high accuracy.

\section{Summary}
In this study, we performed QSVT-based Hamiltonian simulations of the Vlasov-Maxwell equations in plasma physics, conducting a 1D advection test and a 1D1V two-stream instability test using an A100 GPU. For comparison, we also executed Trotter decomposition and classical exact diagonalization.
In the 1D advection test, we discussed the clear differences in the manifestation of numerical errors between the QSVT and Trotter quantum schemes. In the 1D1V two-stream instability test, we demonstrated that phase mixing could be solved numerically stably using QSVT.
The semi-discretized central difference scheme embedded in the quantum framework was found to be a stable scheme that does not cause numerical divergence, as confirmed by von Neumann stability analysis.
Using the error conditions of the QSVT-based Hamiltonian simulation, we presented a specific quantum stability condition corresponding to the CFL condition of classical numerical methods.
Finally, we proposed that using quantum computers for Vlasov-Maxwell simulations could offer advantages not only in terms of increasing the number of grids but also in ensuring numerical stability.
However, as future work, it is necessary to explore new quantum algorithms that incorporate non-Hermitian spatially dependent wind direction evaluation to improve numerical oscillation errors using upwind differences or numerical viscosity.

\section{Acknowledgment}
HH would like to express profound gratitude to Dr. Seiji Zenitani for his invaluable discussions on the two-stream instability, numerical errors, and the CFL condition.
We thank to the Center for Computational Sciences at the University of Tsukuba for providing the computational resources (an A100 GPU) used in this work.
Discussions during the Yukawa Institute for Theoretical Physics (YITP) summer school YITP-W-22-13 on "A novel numerical approach to quantum field theories” were useful as we started this work. 
HH would like to acknowledge the financial support of the Kyushu University Innovator Fellowship Program (Quantum Science Area).
The work of HH and AY is supported by JSPS KAKENHI Grant Numbers\ JP20H01961 and JP22K21345.

\appendix

\section{Implement of Fixed Boundary Condition}\label{apendix:BC}
Embedding fixed boundaries in QSVT-based Hamiltonian simulation is not straightforward. Therefore, we addressed this by adding ancilla qubits and multiple CNOT gates. First, the fixed boundary condition $D_{\mathrm{per}}^{\prime}$ can be written similarly to Eq.~(\ref{eq:D_per}) as
\begin{align}
    D_{\mathrm{per}}^{\prime} = \begin{pmatrix}
    0 & 1 & 0 & \cdots & 0 & 0 & 0 \\
    -1 & 0 & 1 & \cdots & 0 & 0 & 0 \\
    0 & -1 & 0 & \cdots & 0 & 0 & 0 \\
    \vdots & \vdots & \vdots & \ddots & \vdots & \vdots & \vdots \\
    0 & 0 & 0 & \cdots & 0 & 1 & 0 \\
    0 & 0 & 0 & \cdots & -1 & 0 & 1 \\
    0 & 0 & 0 & \cdots & 0 & -1 & 0
    \end{pmatrix}.
\end{align}
To perform this operation, it is necessary to modify the shift operators $U_{\text{Incr.}}$ and $U_{\text{Decr.}}$ in the block-encoded quantum circuit of the effective Hamiltonian. We explain this using the quantum circuit for the 1D Vlasov equation as an example, as shown in Fig.~\ref{fig:1DHamiltonian}.
\begin{align}
\raisebox{12mm}{ 
  \Qcircuit @C=1em @R=1.2em {
  &\lstick{\tanc_3\hspace{1mm} }& \qw &\ctrlo{2}&\ctrl{2} & \qw & \qw & \qw\\
  &\lstick{\tanc_4\hspace{1mm} }& \qw & \qw & \qw & \targ & \targ & \qw \\
  &\lstick{\tx\hspace{1mm} }& {/} \qw & \gate{U_{\text{Incr.}}} & \gate{U_{\text{Decr.}}} & \ctrlo{-1} & \ctrl{-1} & \qw
  }}.
\end{align}
By adding an ancilla qubit $\tanc_4$ for removal and using CNOT gates to flip the ancilla qubit state that holds the boundary value of $\tx$, we effectively remove the boundary value.

\bibliographystyle{jpp}

\bibliography{jpp-instructions}

\end{document}